\documentclass{kapproc} 

\kluwerbib

\begin{document}

\articletitle{Magnetars and pulsars: a missing link}


\author{Bing Zhang}
\affil{Department of Astronomy \& Astrophysics \\
Pennsylvania State University}
\email{bzhang@astro.psu.edu}


\begin{abstract}
There is growing evidence that soft gamma-ray repeaters (SGRs) and
anomalous X-ray pulsars (AXPs) are isolated neutron stars with
superstrong magnetic fields, i.e., magnetars, marking them a 
distinguished species from the conventional species of
spindown-powered isolated neutron stars, i.e., radio pulsars. The
current arguments in favor of the magnetar interpretation of SGR/AXP 
phenomenology will be outlined, and the two energy sources in
magnetars, i.e. a magnetic dissipation energy and a spindown energy,
will be reviewed. I will then discuss a missing link between magnetars
and pulsars, i.e., lack of the observational evidence of the
spindown-powered behaviors in known magnetars. Some recent theoretical 
efforts in studying such behaviors will be reviewed along with some
predictions testable in the near future.  
\end{abstract}

\begin{keywords}
pulsars, magnetars, neutron stars, magnetic fields
\end{keywords}

\section{Soft gamma-ray repeaters and anomalous X-ray pulsars as
magnetars} 
For a long time, radio pulsars have been regarded as the only
manifestation of isolated neutron stars\footnote{The internal
compositions and equations-of-state of ``neutron stars'' are not
well determined. These stars could be in principle more exotic, e.g.,
could be composed of pure strange quark matter (e.g. Xu 2002, in these
proceedings). Here I refer to ``neutron stars'' as a broader class of
objects that includes more exotic categories.}. Recent observational 
developments indicate that isolated neutron stars also manifest
themselves as other species (Pavlov et al. 2002, for a review), among
which soft gamma-ray repeaters (SGRs) and anomalous X-ray repeaters
(AXPs) have attracted growing attention in the neutron star
community. These two types of objects originate, respectively, from the
anomalous species of two distinct classes of phenomenon, i.e.,
gamma-ray bursts and accreting X-ray pulsars, but share many common
features. Recently, two observational facts finally connect a bridge
between SGRs and AXPs. First, after being quiescent for more than
twenty years, SGR 0526-66 is found to have a steep non-thermal
spectrum in the quiescent state which is similar to the non-bursting
AXPs (Kulkarni et al. 2003). Second, soft, repeating bursts were
recently detected from two AXPs, 1E 1048-5937 (Gavriil, Kaspi \& Woods
2002) and 1E 2259+586 (Kaspi \& Gavriil 2002). These suggest that
SGRs/AXPs belong to a unified class of objects. 

In the literature, there exist essentially four types of models to
interpret SGR/AXP phenomenology. These are, according to the sequence
of popularity, the magnetar model (Duncan \& Thompson 1992;
Paczy\'nski 1992; Thompson \& Duncan 1995, 1996; Thompson, Lyutikov \&
Kulkarni 2002), the accretion model involving fossil disks (Chatterjee
et al. 2000; Alpar 2001; Masden et al. 2001), the models involving
strange quark stars (Alcock et al. 1986; Cheng \& Dai 1998; Zhang, Xu
\& Qiao 2000; Usov 2001), and the models involving magnetic white
dwarfs (Paczy\'nski 1990; Usov 1993). It is fair to say that at the
current stage none of the models can interpret {\em all} SGR/AXP
observations satisfactorily. Nonetheless, the magnetar model has its
merit to interpret most observations under one single hypothesis, i.e.,
SGRs/AXPs are neutron stars with superstrong magnetic fields ($\sim
10^{14}-10^{15}$ G at the surface). Other models either have troubles
to interpret some observations (e.g. the accretion model fails to
account for the super-Eddington SGR bursts) or have to
introduce additional assumptions to account for data (e.g. Zhang 2002a
for a review). Below I will list the solid observational facts of
SGR/AXPs and confront them with the magnetar model.

1. {\bf Timing properties.} Known SGRs/AXPs exclusively have long
periods [$P\sim (5-12)$ s] and large spindown rates [$\dot P \sim
5\times (10^{-13}-10^{-10})$ s/s]. Assuming magnetic braking, this
directly refers to a superstrong surface magnetic fields [$B_s \sim
(10^{14}-10^{15})$ G if these objects are neutron stars. Irregular
spindown may be a common feature of these objects, and is not
necessarily related to the bursting behavior. This could be
accomendated in a magnetar model with twisted magnetosphere (Thompson
et al. 2002). 

2. {\bf Quiescent emission properties.} SGRs/AXPs all display a steady 
luminous X-ray emission with $L_x \sim (10^{35}-10^{36})$ ergs/s,
which could be explained in terms of magnetic dissipation (magnetic
field decay, Thompson \& Duncan 1996; or magnetic enhanced cooling, 
Heyl \& Hernquist 1998; or untwisting of a global current-carrying
magnetosphere, Thompson et al. 2002). Optical/IR counterparts have
been detected from three AXPs (4U 0142+61, 1E 2259+586, and 1E
1048.1-5937), but no promising interpretation within the magnetar
model is proposed. No gamma-ray and radio emission has been firmly
detected from the SGRs/AXPs. 

3. {\bf Burst properties.} SGR bursts are soft and repeating, with 
luminosity ranging from $10^{38}$ ergs/s all the way up to $\sim
10^{45}$ ergs/s (usually super-Eddington, and two most luminous
bursts, namely giant flares, 
have been detected from SGR 0526-66 on March 5, 1979; and from SGR
1900+14 on August 27, 1998). A strength of the magnetar model is
that it can interpret the bursting phenomenology successfully in terms 
of the magnetic cataclysmic dissipation events in superstrong magnetic
fields. Super-Eddington bursts are natural in strong fields in which
the Thomson cross section is suppressed.

4. {\bf Environmental effects.} Most SGRs/AXPs are located close to
supernova remnants (SNRs) in projection. Solid associations with the
SNRs are yet firmly established. Real associations are consistent with
the magnetar theory which predicts that these objects are young neutron
stars, but the SNR ages are not fully consistent with the spindown age 
of these objects. Assuming associations, SGRs have larger proper
motions than AXPs. That one AXP with SNR association, 1E 2259+586,
recently displayed hundreds of repeating bursts make the issue more
complicated. The claim that SGRs/AXPs are born in dense environments
(Marsden et al. 2001) is not confirmed (Gaensler et al. 2001). 

5. {\bf Cyclotron features}. Cyclotron features have been detected in
SGR outbursts (Ibrahim et al. 2002), which is consistent with the
magnetar model if the features are of proton-origin, but refers to a
much lower magnetic field if the features are of electron-origin.

In summary, though not fully unquestionable, the magnetar model is
successful in many respects in interpreting the data. However, there
is hitherto no definite proof that 
SGRs/AXPs are isolated neutron stars powered by superstrong magnetic
fields. I believe that the key to prove the magnetar interpretation
would be looking for a missing link between magnetars and normal
pulsars, which I will lay out in the next section.

\section{Two energy sources in magnetars, and a missing link between 
magnetars and pulsars}

If SGRs/AXPs are magnetars, there should be two independent energy
sources in these objects, i.e., the magnetic energy and the spin
energy of a neutron star. Assuming a dipole geometry, the total
magnetic energy in a magnetar magnetosphere is $E_B \simeq (1/12)
B_p^2 R^3$. Taking $B_p = 6.4\times 10^{19}~{\rm G}~ \sqrt{P \dot P}$,
and $R= 10^6~{\rm cm}~ R_6$, the magnetic energy can be estimated
\begin{equation}
E_B=1.7 \times 10^{46}~{\rm ergs}~(P/5~{\rm s}) \dot P_{-11} R_6^3,
\end{equation}
where $\dot P_{-11}=\dot P/(10^{-11})$. The rotation energy of the
magnetar is
\begin{equation}
E_R=(1/2) I \Omega^2=7.9 \times 10^{44}~{\rm erg}~ I_{45} (P/5~{\rm
s})^{-2}, 
\end{equation}
where $I=10^{45} ~{\rm g~cm^3}~ I_{45}$ is the typical momentum of inertia
of the magnetar. The critical line in the $P-\dot P$ diagram for the
magnetic energy domination is
\begin{equation}
\dot P_{-11} > 5.8 P^{-3} I_{45} R_6^{-3}.
\end{equation}
In reality, what is more relevant is to compare the energy release
{\em rate} of the magnetic energy and the spin energy. The former 
could be in principle written $L_B=d E_B/dt=-(1/6) (d B_p/dt) B_p
R^3$. Theoretically, $d B_p/dt$ is rather uncertain. It is more
straightforward to take $L_B \sim 10^{35}-10^{36}~ {\rm
erg~s^{-1}}$ directly from the observations, e.g. 
\begin{equation}
L_B = 10^{35}~ {\rm erg~s^{-1}~} L_{B,35}(B),
\end{equation}
where $L_{B,35}(B)$ is an unknown function of $B$, but may be
insensitive to $B$ when $B_p \sim 10^{14}-10^{15}$ G. The spindown
luminosity is 
\begin{equation}
L_{sd} = -I \Omega \dot \Omega=4 \pi^2 I P^{-3} \dot P=3.2 \times 10^{33}
~ {\rm erg~s^{-1}~} I_{45}(P/5~{\rm s})^{-3} \dot P_{-11}.
\end{equation}
Let $L_B>L_{sd}$, the condition of magnetic luminosity domination is
\begin{equation}
P > 1.6~{\rm s}~ \dot P_{-11}^{1/3} I_{45}^{1/3} L_{B,35}^{-1/3}(B).
\end{equation}
It is found that for the typical values of $P$ and $\dot P$ of
magnetars, these objects all lie in the magnetic-dominated
regime. Nonetheless, they are not far from the transition boundary. 
More important, all magnetars ought to be born with millisecond
initial period to ensure vigorous dynamo process to occur (Thompson \& 
Duncan 1993), which means that over the early lifetime of a magnetar,
the spindown energy should be the dominant energy source. Even at the
present epoch (for typical $P$ and $\dot P$ of 
magnetars), the spindown luminosities (which marks the magnitudes of
the pulsar behaviors) are not too low. In fact, many pulsars with such 
similar $L_{sd}$'s are detected to be active.

Then there comes a missing link between the magnetars and the radio
pulsars. These two types of isolated neutron stars seem to solely
manifest the two types of energy sources, respectively. The spindown
energy is clearly manifested in pulsars in terms of coherent radio
emission, and non-thermal gamma-ray and X-ray emission; while in
magnetars the magnetic dissipation energy is manifested in the form of 
luminous X-rays in the quiescent state and of soft gamma-rays in the
burst state. Within the dominant energy output channel 
for the spindown luminosity, i.e., the radio band and the gamma-ray
band, magnetars are not firmly detected. If lack of
magnetic-dominated behavior in normal pulsars is understandable
because of their weak fields involved, non-detection of the
spindown-powered behavior in magnetars is in principle not
justified. It is worth emphasizing that lack of radio and gamma-ray 
emission is the prediction of the accretion model for AXPs. Therefore
studying the spindown-powered behavior from magnetars is of great
theoretical and observational interests. Only when any
spindown-powered behavior is firmly detected in SGRs/AXPs, could the
accretion model be completely ruled out, and hence, presenting a final
proof of the magnetar interpretation. 

\section{Spindown-powered activity in magnetars}

The pulsar behavior is marked by the pair-production activity in the 
magnetosphere. Particles are believed to be accelerated in gaps either 
in the polar cap region near the surface (Ruderman \& Sutherland 1975; 
Arons \& Scharlemann 1979; Harding \& Muslimov 1998) or above the null 
charge surface (Cheng, Ho \& Ruderman 1986). Accelerated primary
particles radiate through curvature radiation or inverse Compton
scattering, and the resultant gamma-rays produce electron-positron
pairs either through one photon ($\gamma(B) \rightarrow e^{+}e^{-}
(B)$) or two photon ($\gamma\gamma \rightarrow e^{+}e^{-}$)
processes. In the polar cap region, the secondary pairs also radiate
via synchrotron radiation and inverse Compton scattering, leading to a 
photon-pair cascade (Daugherty \& Harding 1996; Zhang \& Harding
2000a). The condition that pair production is prohibited defines
radio pulsar death. Conventionally, this is defined through an energy
budget criterion that requires a minimum potential to accelerate
particles to a high enough energy in order to allow pair production to
occur. This defines a pulsar death valley in the long $P$ regime
(e.g. Zhang 2002b for a review). According to this criterion, the
known magnetars are well above the death line, so that their
spindown-powered activity is in principle not prohibited.

In order to interpret the apparent radio quiescence of SGRs/AXPs,
Baring \& Harding (1998, 2001) argued that pair production is
suppressed in magnetars by another more exotic QED process, i.e.,
magnetic photon splitting. This interpretation relies on the
assumption that all three photon splitting modes permitted by
charge-parity invariance operate together due to (possible) strong
vacuum dispersion effect in superstrong magnetic fields, so that
photons with both $\perp$ and $\parallel$ polarization modes can
split. In such a case, for a high enough magnetic field strength, photon 
splitting will overwhelm magnetic one photon pair production, so that
gamma-rays essentially split to photons with lower energies before
being materialized, and the magnetar magnetosphere is essentially pair 
free. Zhang (2001) later found that even if one photon pair production 
can be completely suppressed by photon splitting (as conjectured by
Baring \& Harding), pairs may be formed via two-photon pair
production, essentially because the magnetar near surface region is 
a hot environment with a copious soft photon bath generated from
magnetic dissipation. Another issue is that, as long as particles can
keep being accelerated to higher altitudes where magnetic field
strength is considerably degraded, one photon pair production will
overtake photon splitting. This operates for the case of a inner gap
type invoking space-charge-limited flow (Zhang \& Harding 2000b).
Both arguments suggest that a magnetar magnetosphere may not be pair
free. 

Now that the magnetar magnetospheric activity does not differ from 
that of radio pulsars intrinsically, there are good reasons to expect
pulsar-like spin-powered activities from magnetars.  

{\bf 1. Low frequency coherent emission from magnetars?} If pairs are
not prohibited in the magnetar magnetosphere, why SGRs/AXPs are silent
in the conventional radio band? There could be several possible
reasons. The most straightforward possibility is that they are
actually radio loud, but the radio beams do not sweep towards us due
to a very narrow beaming angle of a slow rotator (Gaensler et
al. 2001). Other possibilities include that the typical coherent
emission frequency is not in the conventional radio band (Zhang 2001;
Eichler, Gedalin \& Lyubarsky 2002), or that the coherent condition is
fragile and is destroyed in the hot and twisted magnetospheric
environment.

{\bf 2. Non-thermal high energy emission from magnetars?} Non-thermal
high energy emission is expected from both polar cap cascades and/or from
outer gaps in magnetars. In the outer gap scenario, the gamma-ray
luminosities of the magnetars have been recently predicted (Cheng \&
Zhang 2001; Zhang \& Cheng 2002), which are consistent with the current
upper limits on these objects. According to these predictions, some
SGRs/AXPs should be detectable by the next generation gamma-ray
detector, GLAST. In the polar cap scenario, high energy emission is
also expected, but the typical spectrum would be considerbly shifted
to the softer regime due to the large opacities of the gamma-rays (due
to one-photon, two-photon pair production and photon splitting). Also
the beaming angle is correspondingly smaller. More work in this
direction needs to be carried out.

{\bf 3. High energy neutrinos from magnetars?} Zhang, Dai,
M\'esz\'aros \& Waxman (2002) recently discussed another possible
consequence of the magnetar spindown-powered activity. The discussion
is relevant to one half of the magnetar population, i.e., those with
favorable geometry such that positive ions (likely protons or light
nuclei) are accelerated from the polar cap region. For those magnetars 
that rotate rapidly enough, the acceleration potential would be enough
to accelerate protons to the energies above the photonmeson threshold,
so that these protons will interact with the soft photon bath near the 
surface and produce pions and neutrinos. The condition for the
photomeson interaction threshold is $P<(2.4-6.8)~{\rm s}~ B_{p,15}^{1/2}
R_6^{3/2}$. This defines a ``neutrino death valley'' in the magnetar
$P-\dot P$ (or $P-B_p$) space. Four magnetars are found to be within
or slightly below the valley, which means that under favorable
conditions, they are high energy neutrino emitters. Taking into
account pion cooling, the typical neutrino energy is several TeV. For
on-beam detections, SGR 1900+14 and 1E 1048-5937 have substantial
neutrino fluxes, making them interesting targets for the planned large 
area Cherenkov detectors. The whole magnetar population in the
universe adds an interesting contribution to the diffuse high energy
neutrino background and the diffuse gamma-ray background. 

In the above discussions about magnetar gap accelerations, a dipole
configuration is assumed, whilst a magnetar magnetosphere is certainly
non-dipole. More specifically, Thompson et al. (2002) argue that the
SGR/AXP phenomenology is consistent with the hypothesis that the magnetar
magnetosphere is globally twisted. It would be interesting to study
the charge-depleted acceleration regions in such a twisted
magnetosphere, both near the polar cap region and in the ``outer gap'' 
region. A careful study in this direction is called for.

\section{Concluding remark}

Current data reveals a missing link between magnetars and pulsars. 
Several theoretical efforts have been made to predict spindown-powered 
activities in magnetars. Connecting this missing link with future
observations would provide a solid proof that SGRs/AXPs are indeed
isolated neutron stars with strong magnetic fields, i.e., magnetars.

\acknowledgment{I thank A. K. Harding, P. M\'esz\'aros, Z. G. Dai,
E. Waxman, R. X. Xu and G. J. Qiao for stimulating
collaborations on various topics covered in this review. My research
at Penn State University has been supported by NASA grants NAG 5-9192
and NAG 5-9153.}

\end{document}